\documentclass[a4paper]{article}

\usepackage{INTERSPEECH2018}

\usepackage{multirow}
\usepackage{longtable}
\usepackage{array}
\usepackage{dirtytalk}
\usepackage{dblfloatfix}
\usepackage{subcaption}
\usepackage{authblk}
\usepackage{cite,amsmath,amssymb}
\usepackage{graphicx,color}
\usepackage{multirow}
\usepackage[dvipsnames]{xcolor}

\title{Variational Autoencoders for Learning Latent Representations of Speech Emotion: A Preliminary Study}
\name{Siddique Latif$^1$, Rajib Rana$^2$, Junaid Qadir$^1$, Julien Epps$^3$}
\address{
  $^1$Information Technology University (ITU)-Punjab, Pakistan\\
  $^2$University of Southern Queensland, Australia\\
  $^3$University of New South Wales, Sydney, Australia}
\email{siddique.latif@itu.edu.pk, rajib.rana@usq.edu.au, junaid.qadir@itu.edu.pk, j.epps@unsw.edu.au}

\begin{document}

\maketitle
\begin{abstract}
  Learning the latent representation of data in unsupervised fashion is a very interesting process that provides relevant features for enhancing the performance of a classifier. For speech emotion recognition tasks, generating effective features is crucial.  Currently, handcrafted features are mostly used for speech emotion recognition, however, features learned automatically using deep learning have shown strong success in many problems, especially in image processing.  In particular, deep generative models such as Variational Autoencoders (VAEs) have gained enormous success for generating features for natural images. Inspired by this, we propose VAEs for deriving the latent representation of speech signals and use this representation to classify emotions. To the best of our knowledge, we are the first to propose VAEs for speech emotion classification. Evaluations on the IEMOCAP dataset demonstrate that features learned by VAEs can produce state-of-the-art results for speech emotion classification.
\end{abstract}

\section{Introduction}
Recently speech emotion recognition has received significant attention from both industry and academia. It has various applications in human-computer interaction and analysis of human-human interactions. The speech signal has complex distributions with high variance due to various factors such as speaking style, age, gender, linguistic content, environmental and channel effects, emotional state. Understanding the influence of these factors on the speech signal is a crucial problem
for speech emotion recognition. {Although considerable attempts have focused on handcrafting features to capture these 
factors \cite{eyben2016geneva}, automatic learning of features that are sensitive to emotion 
needs more exploration.}  

Deep generative models are recently becoming immensely popular in the deep learning community due to the fact that unlike discriminative approaches, they try to learn the true distribution of the training data and generate new data points (with some variations). In this paper, we are not focused on generating new data but on capitalising the capacity of generative models to learn the true distribution of the data and hence create powerful features, automatically. The most commonly used and efficient generative models are currently Generative Adversarial Nets (GANs) \cite{goodfellow2014generative} and Variational Autoencoders (VAEs) \cite{kingma2013auto}.
While GANs are optimised for generative tasks, VAEs are probabilistic graphical models which are optimised for latent modelling. We therefore focus on VAEs.
There have been many attempts to model natural images using generative models \cite{radford2015unsupervised,larsen2015autoencoding,denton2015deep}, but only some research has been conducted into learning latent representations of speech generation \cite{blaauw2016modeling,hsu2017learning}, conversion \cite{hsu2016voice}, and speaker identification \cite{villalba2017tied}. {Most importantly, the feasibility of VAEs for speech emotion recognition is largely unexplored.}


 In this paper, we conduct a preliminary study to understand the feasibility of VAE for learning the latent representation of speech emotion. We also investigate the performance of a variant of VAE known as Conditional Variational Autoencoder (CVAE) for learning the latent representation of speech emotion.  To objectively measure the performance of this latent representation, we use Long Short Term Memory (LSTM) to classify speech emotion using the latent representation as features. This simultaneously offers the opportunity to validate the performance of VAE for learning latent representation, and delivers a new VAE-LSTM classification framework. Given that Autoencoders (AE) have been widely used for speech emotion, we implement an AE-LSTM model to compare its classification performance with VAE-LSTM. We also compare its classification performance of VAE-LSTM with the recent results in the literature. Our comparisons show that latent representation learned by VAE and its variant CVAE (For brevity we often use the term ``VAEs'' to represent the pair.) can help achieve state-of-the-art speech emotion classification performance. 



\section{Related Work}
\label{RW}
Autoencoders have been extensively used for emotion recognition (e.g.,~\cite{ deng2013sparse, deng2018semisupervised}), however to date, Variational Autoencoders have mainly been used for natural image generation (e.g., \cite{hou2017deep,sonderby2016ladder}). Use of VAEs for speech processing and recognition is very limited. In the speech and audio domain, VAEs have mainly been used for speech generation and transformation \cite{hsu2017learning}. They have also been used to learn phonetic content or speaker identity in speech segments without supervisory data \cite{hsu2017learning,blaauw2016modeling}. Moreover, a framework based on VAE was used in \cite{tan2016learning} to learn both frame-level and utterance-level robust representations. The authors used these salient features along with the other speech features for robust speech recognition. \textcolor{black}{Hsu et al. \cite{hsu2016voice} proposed a VAE based framework for modelling of spectral conversion with unaligned corpora. In this study, the encoder learned the phonetic representation for the speaker, and the decoder reconstructed the designated speaker by removing the demand of parallel corpora for the model training on spectral conversion.} \textcolor{black}{Finally, Blaauw et al. \cite{blaauw2016modeling} used a fully-connected VAE to model the frame-level spectral envelopes of the speech signal. Based on their experiments, the authors found that VAE can achieve similar or comparatively better reconstruction errors than related competitive models like the Restricted Boltzmann machine (RBM).} 

Many researchers have used LSTMs for speech emotion recognition (e.g., \cite{chernykh2017emotion,lee2015high}). In many scenarios, LSTMs are more effective than conventionally-employed support vector machines 
\cite{tian2015emotion}. 
Researchers have also used LSTM networks on the IEMOCAP speech corpus and have shown that \textcolor{black}{ they perform better than powerful methods like Hidden Markov Models 
\cite{wollmer2012analyzing,wollmer2010context}.} Chernykh et al. \cite{chernykh2017emotion} used a Connectionist Temporal Classification (CTC) loss function with LSTM networks for emotion classification, and evaluated it on the IEMOCAP dataset. 
In her worth mentioning work~\cite{gideon2017progressive}, Emily et al. also employed the IEMOCAP database for speech emotion recognition. However, the authors have used transfer learning to leverage information from another database to improve the speech emotion accuracy. Transfer learning is out of the scope of this paper, but in future we would investigate if transfer learning can further enhance the accuracy achieved by our approach. 

\section{Methods}
\subsection{Generating Speech Features using VAE}
Variational Autoencoder (VAE) is a combination of Graphical Models and Neural Networks. It has a similar structure as an Autoencoder (AE) but functions differently. An AE learns a compressed representation of the input and then reconstructs the input from the compressed representation. 
On the other hand, VAE learns the {parameters of a probability distribution} representing the input in a latent space. This is done by making the latent distribution as close as possible to a ``prior''
on the latent variable. The key advantages of the VAE over an AE is that the ``prior'' allows the injection of domain knowledge, enabling estimation of the uncertainty in the prediction, and making it more suitable for speech emotion recognition.

Formally speaking, given any emotion data $X$ the aim of VAE is to find the probability of $X$ with respect to its latent representation $z$:
\begin{equation}
    P(X) = \int {P(X|z)P(z) dz}.
\end{equation}
However, the quantities $P(X|z)$ and $P(z)$ both are unknown. The idea of VAE is to infer $P(z)$ using $P(z|X)$, where $P(z|X)$ is determined using Variational Inference (VI). In VI, $P(z|X)$ is inferred upon minimising the divergence with a known distribution $Q(z|X)$. It becomes~\cite{kingma2013auto},



\begin{eqnarray}
\label{KL1}
    \log{P(X)} = - \{|X-\hat{X}|^2 + KL[Q(z|X)||P(z)]\}
    \end{eqnarray}
As can be seen in ~\eqref{KL1}, the aim of VI is to eventually reduce the reconstruction error and to train the encoder $Q(z|X)$ in such a way that it produces the parameters of the probability distribution for the latent space $z$ based on a known distribution of choice. This will minimise the divergence between $Q(z|X)$ and $P(z)$. For example, if we assume that the latent space will have a normal distribution, we need to train the encoder to generate the mean and covariance. Samples of $P(z|X)$ will be generated using these parameters, which the decoder will use to generate the approximation of $X$.  %


\noindent{\bf Conditional Variational Autoencoder (CVAE)}:
In conventional VAE there is no way to generate specific data, for example a picture of an elephant, if the user inputs an elephant image. This is because the VAE models the latent variable and image directly. To eliminate this problem, the Conditional Variational Autoencoder (CVAE) models both latent variables and the emotion data conditioned on some random variables, $c$. The encoder is therefore conditioned on two variables $X$ and $c$: $Q(z|X,c)$ and the decoder is also conditioned to two variables, $z$ and $c$: $P(X|z,c)$.
There are many possibilities for the conditional variable: it could have a categorical distribution expressing the label, or even could have the same distribution as the data.

{Despite the capabilities of VAE, we are not particularly interested  in generating speech emotion $\hat{X}$. However, when the distance ($|X-\hat{X}|^2$) between the original and the generated emotion becomes smaller than our predefined threshold, we use the parameters of the probability distribution {$P(z|X)$} 
as the features for emotion $X$.} {For imposing conditions on the $P(z|X)$ (i.e. to emulate CVAE), we simply concatenate the speech frame representation in LogMel for any particular emotion $X$ with its emotion class label ($c$) and pass this into the encoder.} 

\subsection{Speech Emotion Classification using LSTM}
LSTM can model a long range of contexts due to the presence of a special structure called the memory cell.
\textcolor{black}{Emotions in speech are context-dependent, therefore the ability to model contextual information makes LSTM suitable for 
speech emotion recognition
\cite{tian2016recognizing}. }

The LSTM memory cell is built into a memory block, which constitutes the hidden layers of LSTM. There are three “gate” units in the memory cell - the input, output, and forget gate, which are used to perform reading, writing, and resetting of information, respectively. When the feature representations from the VAE are input to the LSTM, the input gate enables a memory block to selectively control the incoming information and store in the internal memory. The output gate decides what part of the information will be output, and a forget gate selectively clears the speech emotional contents from the memory cell.



To use LSTM for emotion classification, its output vector (end layer) is projected onto a vector with a length of the number of emotion classes. Projection is done using simple functions $Q= Wx$, where $x\in{R}^n$ is the LSTM output vector, $W\in{R}^{m\times n}$ is a weight vector and $Q\in{R}^m$ is the vector having the same length as the number of classes $m$. The vector $Q$ is then mapped onto a probability vector with values in $[0,1]$ having sum of the probabilities equates to $1$. The highest probability indicates the identified class.



\begin{figure}[!ht]
\centering
\centerline{\includegraphics[width=.4\textwidth]{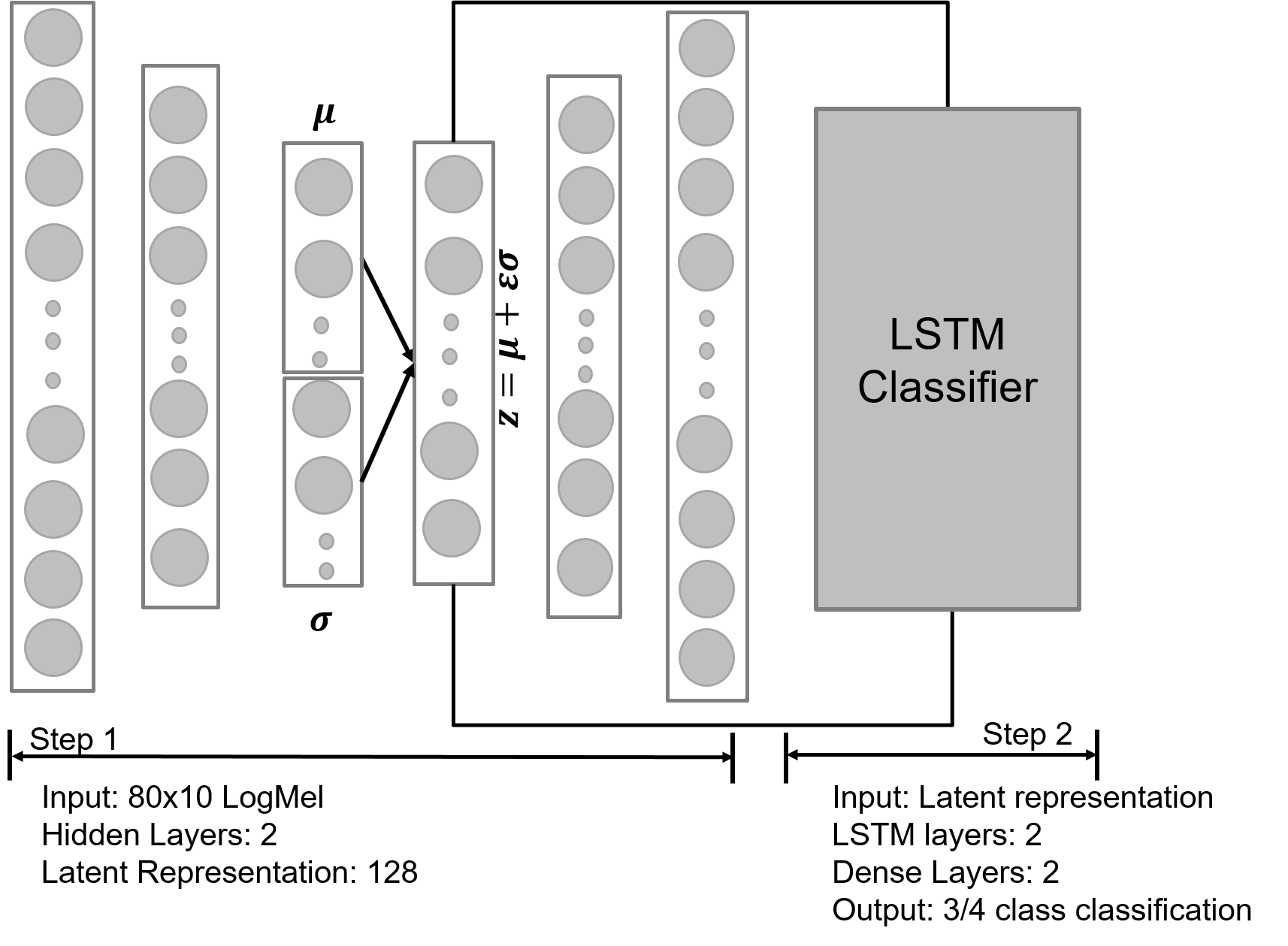}}
\caption{Overall Classification Framework.}
\label{fig:model}
\end{figure}

The overall classification framework  has been shown in Figure~\ref{fig:model}. Previous studies have concluded that the performance of the LSTM model can be enhanced by using more predictive and knowledge-inspired features despite the limited training examples~\cite{tian2016recognizing,tian2015emotion,chao2015long}. Therefore, LSTM is a natural choice for us to use with features generated by VAEs.

\begin{table*}[!b]
\centering
\scriptsize
\caption{Accuracy (\%) comparison amongst different models for categorical classification.}
\label{table: Com}
\begin{tabular}{|c|c|c|c|c|c|c|c|c|c|c|c|}
\hline
\multirow{2}{*}{Data} & \multicolumn{2}{c|}{AE-LSTM} & \multicolumn{2}{c|}{VAE-LSTM} & \multicolumn{2}{c|}{\textbf{CVAE-LSTM}} & \multicolumn{3}{c|}{Attentive CNN \cite{neumann2017attentive} (WA)}     & \multirow{2}{*}{BLSTM~\cite{chernykh2017emotion} (WA)} & \multirow{2}{*}{BLSTM \cite{lee2015high} (WA)} \\ \cline{2-10}
                      & WA            & UA           & WA            & UA            & WA                      & UA            & LogMel & MFCC  & eGeMAPS &                             &                             \\ \hline
Improvised            & 59.84         & 58.32        & 63.21         & 60.91         & \textbf{64.93}          & 62.81         & 61.716      & 61.35     & 61.27        & 54                          & 62.85                       \\ \hline
Scripted              & 52.68         & 48.52        & 53.74         & 52.23         & 55.71                   & 53.50         & 52.64       & 53.19     & 53.19        & NA                          & NA                          \\ \hline
Complete Data         & 58.16         & 55.42        & 60.71         & 56.08         & 61.08                   & 58.10         & 54.86       & 55.12     & 54.78        & NA                          & NA                          \\ \hline
\end{tabular}
\end{table*}

\section{Experimental Setup}
\label{ER}


\subsection{Speech Corpus}

For experimentation, we selected the Interactive Emotional Dyadic Motion Capture (IEMOCAP) \cite{busso2008iemocap} dataset, which is widely used for speech emotion recognition. IEMOCAP is a multimodal corpus containing recordings of ten actors over five sessions. Each session contains one female and one male speaker. The data includes two types of dialogues: scripted and non-scripted. In the non-scripted dialogue, the speakers were instructed to act without pre-written scripts. For the scripted dialogue data, the actors followed a pre-written script. Annotation was performed by 3-4 assessors based on both video and audio streams. Each utterance was annotated using  10 categories: neutral, happiness, sadness, anger, surprise, fear, disgust frustration, excited, and other. {To better compare the results with related work, we computed our results for improvised, scripted and complete data (including both improvised and scripted). We considered four emotions: neutral, happiness, sadness, and anger, by combining happiness and excited as one emotion, following the state-of-the-art studies on this corpus \cite{xia2017multi,neumann2017attentive}.

{IEMOCAP data were also annotated on three continuous dimensions: Arousal (A), Power (P), and Valence (V).} For comparison of our classification results with the state-of-the-art approaches in \cite{tian2016recognizing,tian2015emotion}, we also consider the above emotion dimensions. However, to maintain it as a classification problem, like~\cite{tian2016recognizing,tian2015emotion}, within each dimension we created three categories: low (values less than 3), mid (values equal to 3) and high (values greater than 3).} 


\subsection{Speech Data Processing}
We consider the LogMel speech frame representation, as used in \cite{abdel2014convolutional,neumann2017attentive}.
Again following the above studies, a Hamming window of length 25ms with 10ms frame-shift was applied to the speech signal, and the discrete Fourier transform coefficients were computed. We then computed 80 mel-frequency filter-banks. The feature set was formulated by taking the  logarithmic power of each mel-frequency band energy.

\subsection{Configuration of VAE and LSTM}
We input speech segments of length 100ms into the VAE for latent representation of data. This speech segment of 800 features is represented in a latent space of 128. We used two encoding layers with 512 and 256 hidden units respectively. The number of hidden units were chosen based on intuition from prior work on autoencoders \cite{kingma2013auto} and on speech recognition using VAEs \cite{hsu2017learning}. 

We used the Adam (adaptive moment estimation) optimiser, which is a Stochastic Optimisation Algorithm widely used to update network weights iteratively based on the training data \cite{kingma2014adam}. The values of the various parameters used in the Adam optimiser were as follows: $\beta1$=0.999 and $\beta2$=0.99, $\epsilon$=$10^{-8}$ and learning rate =$10^{-3}$. \textcolor{black}{These values were chosen in an iterative manner to obtain the minimum reconstruction loss of the autoencoder networks}.  
{We used the reparameterization trick~\cite{kingma2013auto} to approximate the latent space $z$ with normally distributed $\delta$ by setting $z=\mu+\delta \odot \sigma$, where $\odot$ denotes element-wise multiplication, $\delta\sim N(0,1)$, and $z\sim N(\mu, \sigma)$. }

In CVAE we conditioned the VAE on the categorical emotion labels. To benchmark the performance of VAE, we also used a conventional autoencoder (AE) having the same architecture (i.e., hidden units, layers and model parameters), except for the Gaussian layer, which was replaced with a fully connected layer.

 \textcolor{black}{Our LSTM model consisted of two consecutive LSTM layers with the activation of the hyperbolic tangent. The hidden states of the second LSTM layer were connected to the dense layer and the outputs of the dense layer were fed into the softmax layer for classification of both categorical and dimensional class labels.} The network parameters were chosen through cross-validation experiments. \textcolor{black}{As a common setup, we used the Adam optimiser \cite{kingma2014adam} with default learning rate of $10^{-3}$ by following \cite{kim2017learning}. To avoid overfitting, we used early stopping criteria with the maximum number of epochs equal to 20}. All the experiments were performed using an Nvidia Quadro M5000 with 8 GB memory.


\section{Results}
The latent representations generated by both VAEs and AE were input to an LSTM network for classification. The segment-level latent representations obtained by autoencoder networks were merged into the whole utterance-level features for classification of emotions as in  \cite{han2014speech,zhao2017recurrent}. 
Because the IEMOCAP corpus did not have a have a split of training and testing data,  we investigated the performance of our model by training it in the speaker-independent manner. This also allowed us to compare our results with previous studies. We adopted a leave-one-session-out cross-validation approach and evaluated the models for both weighted accuracy (WA) and unweighted accuracy (UA) for categorical dimensions. 
{For dimensional annotations, we followed evaluations strategies in \cite{tian2015emotion,tian2016recognizing} to be able to compare with these studies. We report the F-measures scores over the test dataset. The models were trained using 90\% of data and testing was performed on the remaining 10\% of unseen data.} 
\begin{figure*}[!h]%
\centering
\begin{subfigure}{0.72\linewidth}
\includegraphics[width=\linewidth]{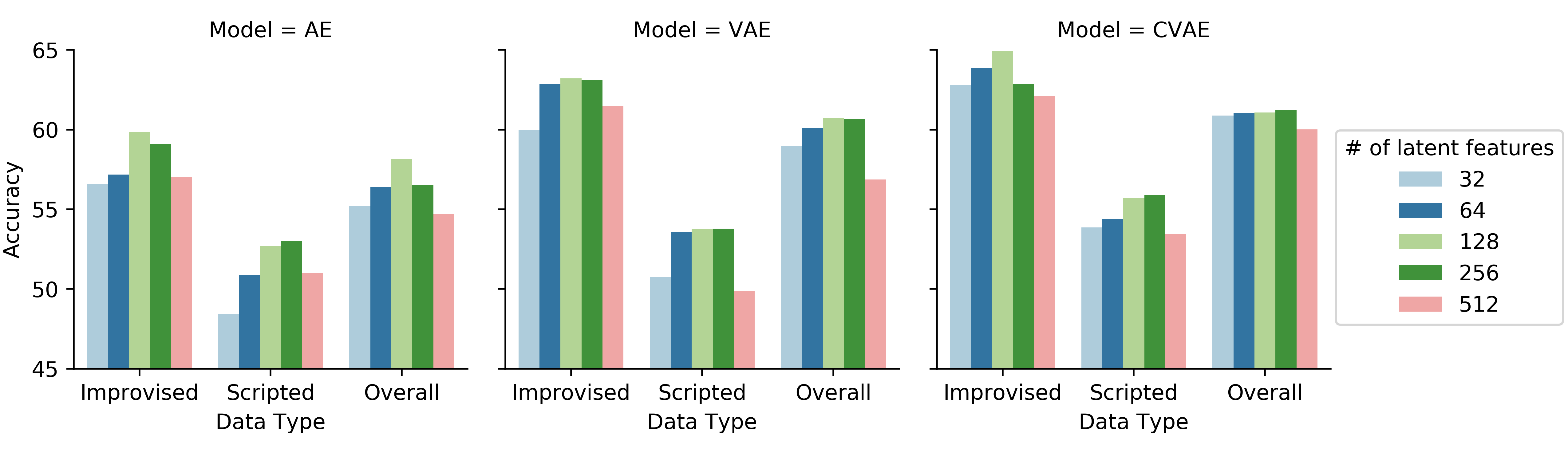}%
\captionsetup{justification=centering}
\caption{} %
\label{fig:cate}%
\end{subfigure}%
\begin{subfigure}{0.28\linewidth}
\includegraphics[width=\linewidth]{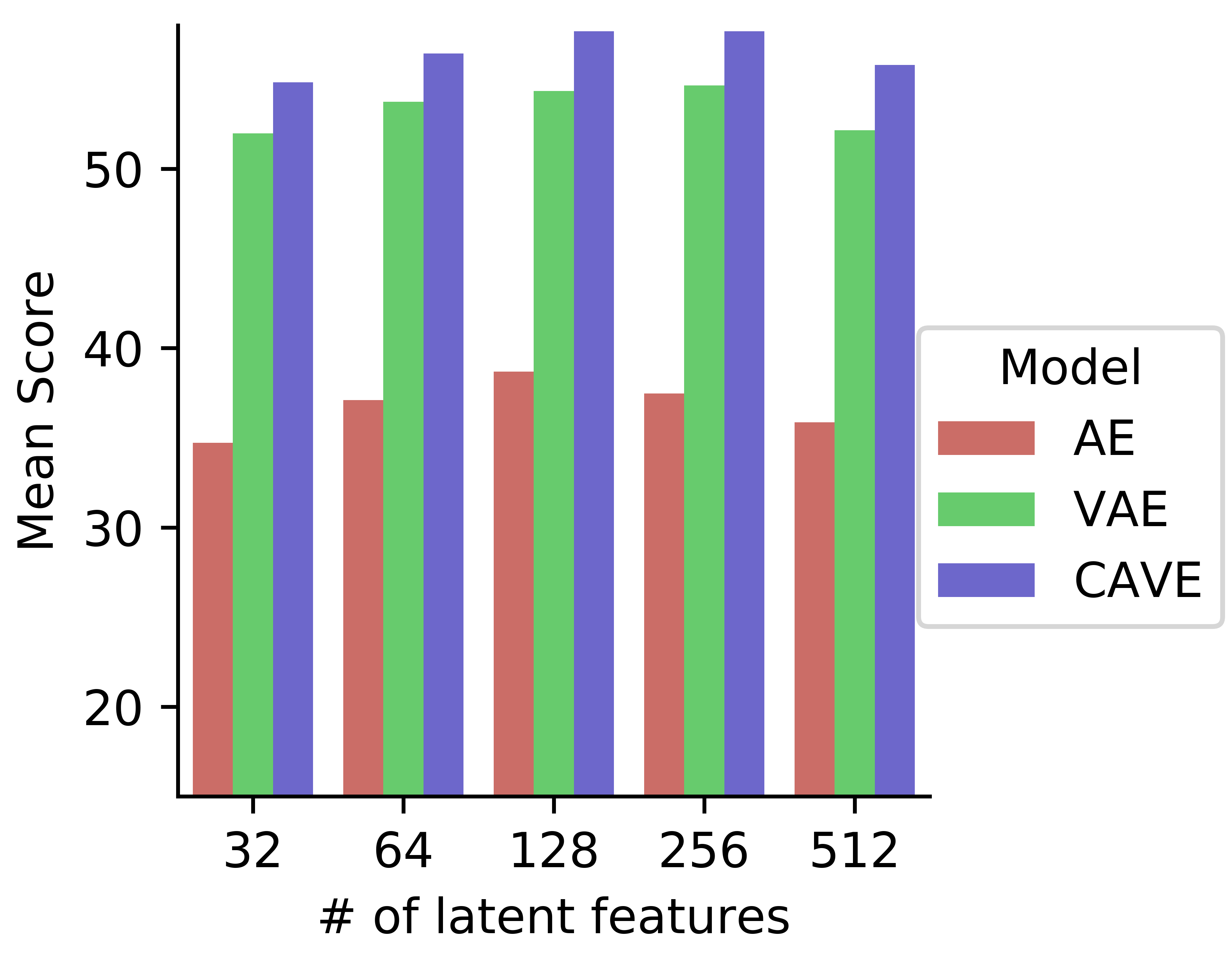}%
\captionsetup{justification=centering}
\caption{}%
\label{fig:dim}%
\end{subfigure}
\caption{{Results using different number of latent features both on categorical and dimensional annotations. Figure \ref{fig:cate} shows the effect of different number of features on categorical classification accuracy and \ref{fig:dim} presents the corresponding trend of mean score for dimensional annotation.}
}
\end{figure*}
\subsection{Classification Performance for Categorical Emotions}
Table \ref{table: Com} shows the five-fold classification results on different subsets of the IEMOCAP data. It can be noted that the features learned by VAE produces better classification performance when compared with the conventional autoencoder. The representations learned by CVAE are highly predictive, which further outperform that learned by VAE. 


{In Table \ref{table: Com}, we also compare different approaches on IEMOCAP used in the literature with our proposed approach. Lee et al. \cite{lee2015high} proposed an extreme learning machine (ELM) based RNN model using bidirectional-LSTM (BLSTM) model and achieved 62.85\% accuracy. The authors used low-level acoustic features and MFCC along with their derivatives, as a feature set to the model. In \cite{neumann2017attentive}, authors used different types of features and evaluated single view (SV) as well as multi-view (MV) attentive CNN on IEMOCAP data using four emotions (as we used). We mention their best results (SV or MV) in the table. Chernykh et al. \cite{chernykh2017emotion} used three different type of features (MFCC, chromagram, and spectrum properties) and report 54\% accuracy using BLSTM. Using CVAE derived features, we achieve 64.93\% accuracy, which is very competitive with respect to the literature.}

\subsection{Classification Performance for Dimensional Emotions}
Table \ref{table: Com2} presents the 10-fold cross-validation results on dimensional annotation using IEMOCAP data, where ``Mean'' represents the arithmetic mean of all three emotional dimensions: Arousal (``A''), Power (``P''), and Valence (``V''). The results are calculated on the basis of classifying the three subcategories: low, mid and high within each emotion dimension.  We compare the performance of our proposed methods with an autoencoder model and also with some recent studies in the literature. Both VAE-LSTM and CVAE-LSTM significantly outperform the AE-LSTM model, while CVAE-LSTM producing the best performance. 
\begin{table}[!h]
\centering
\scriptsize
\caption{Results on IEMOCAP data for dimensional annotations.}
\begin{tabular}{|m{2.2cm}|m{0.7cm}|m{0.7cm}|m{0.7cm} |m{1.2cm}|}
\hline
Method
&A (\%)
&P (\%)
&V (\%)
&\textbf{Mean (\%)}
\\ \hline
\begin{tabular}[c]{@{}l@{}}AE-LSTM\end{tabular}
&\begin{tabular}[c]{@{}l@{}}42.21\end{tabular}
&\begin{tabular}[c]{@{}l@{}}38.25\end{tabular}
&\begin{tabular}[c]{@{}l@{}}35.58\end{tabular}
&\begin{tabular}[c]{@{}l@{}}38.68\end{tabular}
\\\hline
\begin{tabular}[c]{@{}l@{}}VAE-LSTM\end{tabular}
&\begin{tabular}[c]{@{}l@{}}61.35\end{tabular}
&\begin{tabular}[c]{@{}l@{}}53.18\end{tabular}
&\begin{tabular}[c]{@{}l@{}}48.46\end{tabular}
&\begin{tabular}[c]{@{}l@{}}54.33\end{tabular}
\\\hline
 \begin{tabular}[c]{@{}l@{}}{\bf CVAE-LSTM}\end{tabular}
&\begin{tabular}[c]{@{}l@{}}62.73\end{tabular}
&\begin{tabular}[c]{@{}l@{}}53.84\end{tabular}
&\begin{tabular}[c]{@{}l@{}}52.69\end{tabular}
&\begin{tabular}[c]{@{}l@{}}{\bf 56.42}\end{tabular}
\\ \hline
\begin{tabular}[c]{@{}l@{}}DN features~\cite{tian2015emotion}\end{tabular}
&\begin{tabular}[c]{@{}l@{}}41.6\end{tabular}
&\begin{tabular}[c]{@{}l@{}}37.8\end{tabular}
&\begin{tabular}[c]{@{}l@{}}34.0\end{tabular}
&\begin{tabular}[c]{@{}l@{}}37.8\end{tabular}
\\ \hline
\begin{tabular}[c]{@{}l@{}}DN+LLD features~\cite{tian2015emotion}\end{tabular}
&\begin{tabular}[c]{@{}l@{}}53.9\end{tabular}
&\begin{tabular}[c]{@{}l@{}}51.6\end{tabular}
&\begin{tabular}[c]{@{}l@{}}39.5\end{tabular}
&\begin{tabular}[c]{@{}l@{}}48.3\end{tabular}
\\ \hline
\begin{tabular}[c]{@{}l@{}}eGeMAPS~\cite{tian2016recognizing} \end{tabular}
&\begin{tabular}[c]{@{}l@{}}60.1\end{tabular}
&\begin{tabular}[c]{@{}l@{}}52.2\end{tabular}
&\begin{tabular}[c]{@{}l@{}}46.6\end{tabular}
&\begin{tabular}[c]{@{}l@{}}53\end{tabular}
\\ \hline
\begin{tabular}[c]{@{}l@{}}Hierarchical\\Feature Fusion~\cite{tian2016recognizing}\end{tabular}
&\begin{tabular}[c]{@{}l@{}}61.7\end{tabular}
&\begin{tabular}[c]{@{}l@{}}52.8\end{tabular}
&\begin{tabular}[c]{@{}l@{}}51.2\end{tabular}
&\begin{tabular}[c]{@{}l@{}}55.3\end{tabular}
\\\hline
\end{tabular}
\label{table: Com2}
\end{table}

Studies \cite{tian2015emotion,tian2016recognizing} that we have compared with in Table~\ref{table: Com2} used different types of features, such as knowledge-inspired disfluency and nonverbal vocalization (DN) features, and statistical Low-Level Descriptor (LLD) features, as an input to the LSTM model. The highest score they achieve is 55.3\% (Mean score), which we closely outperform using our proposed CVAE-LSTM model (Mean score 56.42\%).

\subsection{Number of Latent Features Versus Accuracy}
In all the results reported above, we have used a latent space size 128, which essentially means we have used 128 set of mean and variances (since, $z=\mu+\delta \odot \sigma$) of a normal distribution as latent features. However, we also investigate the impact of a higher and lower number of latent features.

{Figure \ref{fig:cate} and \ref{fig:dim} show the trend of results using different number of latent features for categorical and dimensional emotions, respectively. Across all of AE, and VAEs, a very small number of features (32) perform poorly. However, a very large number of features (512) does not produce the best performance as well. Within this lower and higher bound, only an insignificant improvement can be observed with the increase of number of features. Based on these results we conclude that a suitable number of latent features needs to be determined empirically to avoid selecting a very small or a very large number of features. 
}

\vspace{0.3cm}

\section{Conclusion}
\label{Con}


In this paper we demonstrate that VAEs can effectively learn latent representation of speech emotion, which offers great potential for learning powerful features, automatically. We show that this helps achieve high classification accuracy when combined with a classifier of natural choice, LSTM, as LSTM has the intrinsic capacity to model contextual information like speech emotion, also an LSTM model can be enhanced by using more predictive and knowledge-inspired features.  We analyse both categorical and dimensional emotions and comparing the emotion classification results with that of a widely used AE-LSTM model, we show that VAEs offer great promise by producing state-of-the-art results. We also analyse the impact of the number of latent features on classification accuracy with a view to determining the optimal number of features. However, we conclude that the suitable number of features needs to determined empirically. Overall, the preliminary results presented in this paper demonstrate that it is highly feasible to automatically learn features for speech emotion classification using deep learning techniques, which  will potentially motivate researchers to further innovate in this space.

\section{Acknowledgements}
This research is partly supported by Advance Queensland Research Fellowship, reference AQRF05616-17RD2. 




\end{document}